\documentclass{bolognino}

\def\Journal#1#2#3#4{{#1} {\bf #2}, #3 (#4)}

\def\NIMA{{\em Nucl. Instrum. Methods} A}

\def\PRL{\em Phys. Rev. Lett.}

\def\RMP{\em Rev. Mod. Phys.}
\def\EPJ{{\em European. Physical. Journal.}C}
\def\EUPJ{{\em Eur. Phys.} J}
\def\RMP{\em  Rev. Mod. Phys.}

\def\be{\begin{equation}}
\def\ee{\end{equation}}
\def\bea{\begin{eqnarray}}
\def\eea{\end{eqnarray}}

\newcommand{\Photo}{}

\begin{document}
\vspace*{4cm}
\title{DIRECT SEARCH OF DARK MATTER WITH THE SABRE EXPERIMENT}

\author{ IRENE BOLOGNINO on behalf of the SABRE Collaboration}

\address{Department of Physics, Universit\`{a} degli Studi di Milano, INFN Sezione di Milano\\
Via Celoria 16, Milano, Italy}

\maketitle\abstracts{
The interaction rate of hypothesized dark matter particles in an Earth-bound detector is expected to undergo an annual modulation due to the planet's orbital motion. The DAMA experiment has observed such a modulation with high significance in an array of scintillating NaI(Tl) crystals. This claim is still unverified inasmuch as the other expe\-ri\-ments involved in this research use different dark matter targets and cannot be compared with DAMA in a model-independent way. The SABRE experiment seeks to provide a much-needed model-independent test by developing highly pure NaI(Tl) crystal detectors with very low ra\-dio\-acti\-vi\-ty and deploying them into an active veto detector that can reject key backgrounds in a dark matter measurement. The final layout of SABRE will consist of a pair of twin detectors at LNGS (Laboratori Nazionali del Gran Sasso, Italy) and SUPL (Stawell Underground Physics Laboratory, Australia). The combined analysis of data sets from the two hemispheres will allow to identify any terrestrial contribution to the modulating signal. 
This article gives an overview of the detector design together with the results of Monte Carlo simulations and of the status of SABRE proof-of-principle activities at LNGS.}

\section{Motivation}

One of the fundamental techniques in direct search for Dark Matter (DM) is the observation of the annual modulation, a strong model independent signature. Hypothesized DM particles, such as Weakly Interacting Massive Particles (WIMPs), are expected to interact with detector target nuclei resulting in nuclear recoil energy being released. The event rate would be expected to show a sinusoidal behavior due to the Earth's motion through the  WIMP halo \cite{modulation}. The DAMA experiment (short for DAMA/NaI and its upgrade DAMA/LIBRA), located at Laboratori Nazionali del Gran Sasso (LNGS) in Italy, has been measuring an annual modulation in NaI(Tl) crystals for almost two decades with high statistical significance \cite{dama}. 
This claim is still unverified inasmuch as other ex\-pe\-ri\-ments looking for DM such as LUX \cite{LUX}, SuperCDMS \cite{CDMS} and XENON \cite{XENON} are based on different target materials so that their results cannot be directly compared with DAMA. A model-independent test can be performed with SABRE (Sodium-iodide with Active Background REjection).

\section{The SABRE experiment}\label{sec:sabrexp}
\begin{figure} [h!]
\begin{minipage}{0.6\linewidth}
\centerline{\includegraphics[width=.8\linewidth]{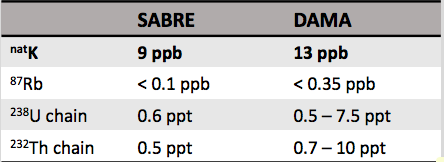}}
\end{minipage}
\hfill
\begin{minipage}{0.3\linewidth}
\centerline{\includegraphics[width=.7\linewidth]{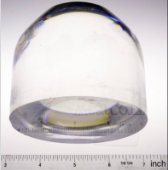}}
\end{minipage}
\hfill
\caption[]{(Left) Results of plasma mass spectroscopy analysis for the most relevant intrinsic impurities in the SABRE 2-kg test crystal \cite{sabreppb} in comparison with DAMA 10-kg crystals \cite{damappb}. (Right) Picture of the 2-kg NaI(Tl) crystal.}
\label{fig:cristallo}
\end{figure}

The SABRE collaboration includes $\sim$ 50 people spread in 10 institutions in Italy, US, and Au\-stra\-lia.
The strategy to reach high sensitivity and test DAMA's results is based on the following four pillars:\\
\begin{enumerate}
\item{The development of high purity crystals starting from ultra-high purity powder.}
\item{The use of an active veto and passive shielding for efficient background rejection.}
\item{The choice of high quantum efficiency photomultipliers (PMTs) with low radioactivity. This allows a direct coupling with the crystal without requiring light guides.}
\item{The design based on two twin detectors located both in the Northern hemisphere (LNGS, Italy) and in the Southern hemisphere (Stawell Underground Physics Laboratory, Victoria, Australia). The combined analysis will reduce seasonal systematic effects.}
\end{enumerate} 
A crucial element for SABRE is the development of extremely pure crystals. Ultra-pure powder has been produced by Sigma Aldrich company. The growth of a 2-kg crystal showed a high purity level of $\sim$9 ppb of \textsuperscript{nat}K (Fig. \ref{fig:cristallo}). Currently a crystal of final size (about 5 kg) is being grown. \\

Each crystal will be coupled to two PMTs, encased in a highly pure copper enclosure and submerged in 2 tons of liquid scintillator acting as a veto.  The most dangerous isotopes in the Region Of Interest (ROI) 2-6 keV  are \textsuperscript{40}K and cosmogenic \textsuperscript{22}Na. \textsuperscript{40}K decays via electronic capture in an excited state of \textsuperscript{40}Ar with a branching ratio (BR) of 11\% emitting a 1.46 MeV $\gamma$. The hole in the K-shell gives an Auger e\textsuperscript{-} or an X-ray of 3.2 keV. The similar decay scheme of \textsuperscript{22}Na gives a 0.8 keV X-ray and a 1.2 MeV $\gamma$ (BR $\sim$ 10\%).  By catching these high energy gammas in the active veto surrounding the crystal, both these background components can be efficiently reduced in SABRE. 
From Monte Carlo simulations, the veto efficiency in rejecting  \textsuperscript{40}K events is $\sim$ 84\%. \\
The SABRE experiment is planned in two phases. The ``Proof-of-Principle'' (PoP) phase is about to start at LNGS and consists of the deployment of one 5-kg NaI(Tl) crystal in the active veto. The goal is the validation of the overall experimental strategy, and the complete characterization of the crystal background and of the veto efficiency. The final phase of the experiment will consist in two $\sim$50 kg detectors located in both hemispheres. The Southern site is under implementation and it is expected to be completed by the beginning of 2019.

\section{The Proof-of-Principle (PoP) Phase}\label{sec:pop}

\begin{figure} [h!]
\begin{minipage}{0.25\linewidth}
\centerline{\includegraphics[width=.9\linewidth]{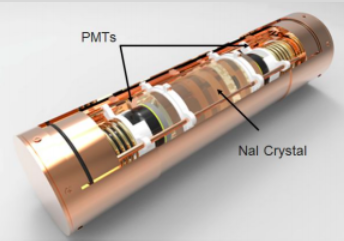}}
\end{minipage}
\hfill
\begin{minipage}{0.4\linewidth}
\centerline{\includegraphics[width=1.1\linewidth]{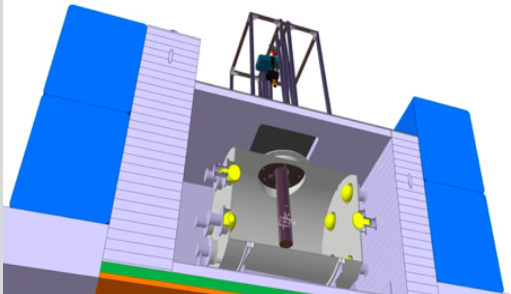}}
\end{minipage}
\hfill
\begin{minipage}{0.3\linewidth}
\centerline{\includegraphics[width=1.\linewidth]{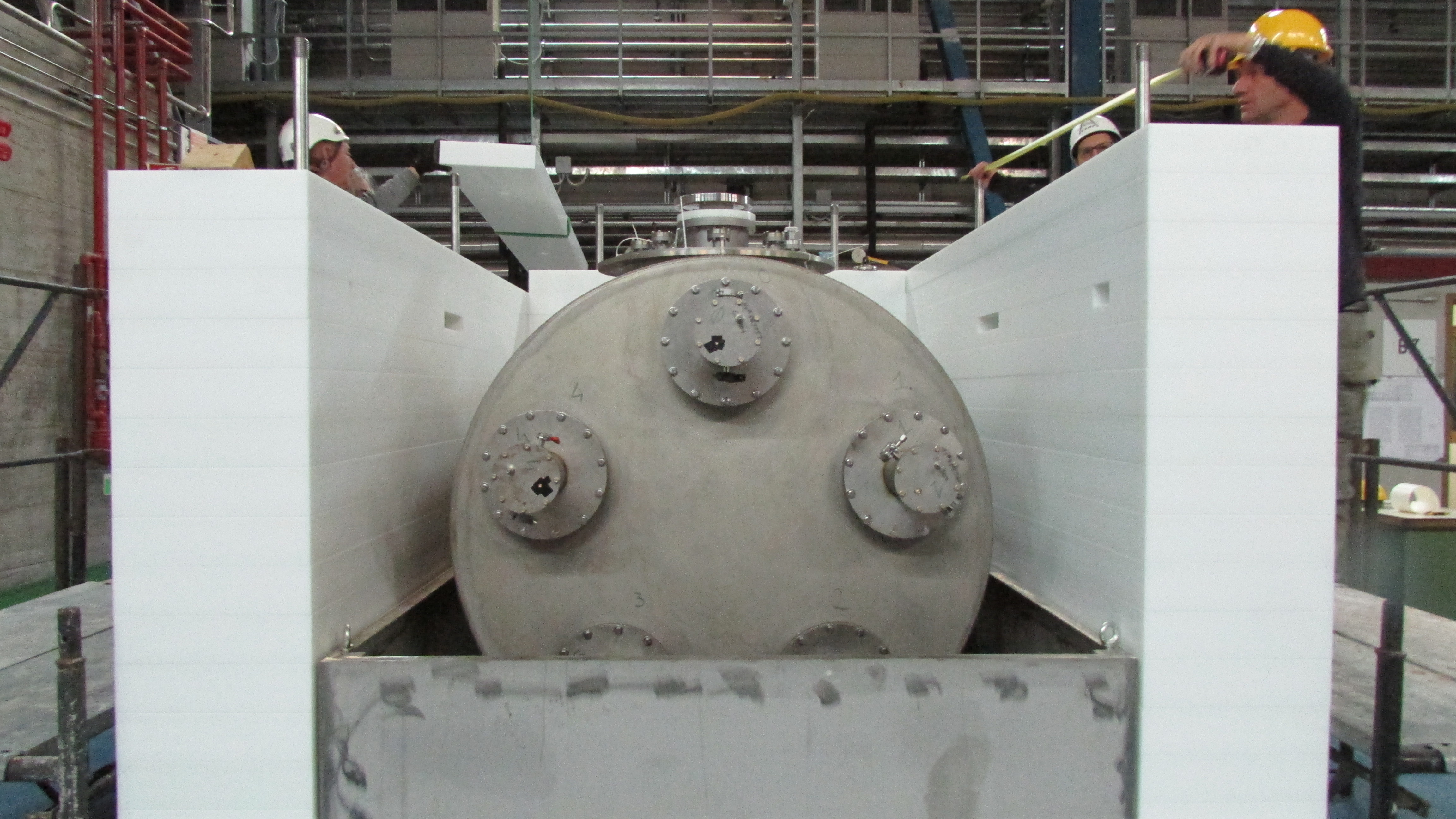}}
\end{minipage}
\hfill
\caption[]{(Left) Copper enclosure where the crystal coupled with 2 PMTs will be located. (Center) Cross-section of the SABRE PoP showing the enclosure, the veto PMTs, the vessel, the external shielding and the crystal insertion system to insert the enclosure in the vessel. (Right) Picture of the veto vessel inside the partially-installed external shielding.}
\label{fig:PoP}
\end{figure}

The SABRE PoP is currently being completed in the hall C at LNGS and it is made of the following components from inward to outward (Fig. \ref{fig:PoP}):
\begin{itemize}
\item{A 5-kg NaI(Tl) crystal directly coupled to two 3" Hamamatsu R11065-20 PMTs with high quantum efficiency (QE) and low radioactivity and encased in a highly pure copper  enclosure.}
\item{The enclosure is connected to a steel bar and inserted into the vessel through a copper tube, that keeps the enclosure insulated from the liquid scintillator. The volume inside the copper tube will be continuously flushed with nitrogen to suppress radon contamination.}
\item{A veto composed of a stainless steel vessel with dimensions 1.4 m diameter and 1.5 m length filled with two tons of liquid scintillator (Pseudocumene+3 g/l of PPO). The vessel's inner surfaces are covered with lumirror in order to improve light reflectance. Scintillation light from the veto is observed by ten 8" Hamamatsu R5912 PMTs with high QE.}
\item{A composite passive shielding made of polyethylene (walls thickness=40-65 cm, top layer thickness=10 cm, floor thickness=10 cm) and water tanks (top thickness=80 cm, sides thickness=90 cm) to suppress external backgrounds. Additionally, a 2 cm steel plate is placed on the top and a layer of 15 cm lead on the floor, under the polyethylene. }
\end{itemize}    

\subsection{Background simulations}\label{subsec:mc}

\begin{figure} [h!]
\begin{minipage}{0.5\linewidth}
\centerline{\includegraphics[width=1.\linewidth]{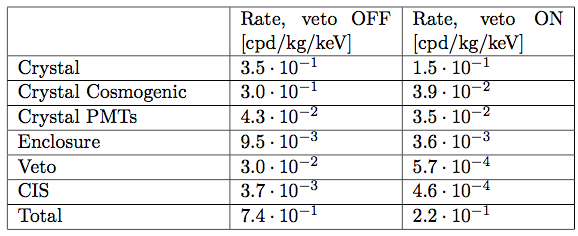}}
\end{minipage}
\hfill
\begin{minipage}{0.5\linewidth}
\centerline{\includegraphics[width=.9\linewidth]{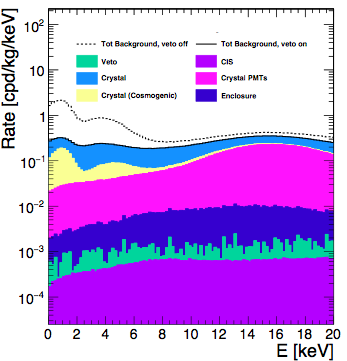}}
\end{minipage}
\hfill
\caption[]{(Left)  Background  from each detector element without and with the veto rejection.  (Right)  Stacked backgrounds and the overall background with veto on (solid black line) and veto off (dashed black line).}
\label{fig:mc}
\end{figure}

The background has been fully estimated using Monte Carlo simulations in the 2-6 keV ROI. The activity levels of the setup components were taken from measurements performed on SABRE materials, when available, or from screening campaigns conducted by other experiments  \cite{damappb} \cite{xenon} \cite{anais}. The results are reported in Fig. \ref{fig:mc} where it is possible to notice how the background in the low energy region is mainly due to crystal residual radioactivity, while the other sources are below this contribution. Crystal cosmogenic activation is calculated after 180 days underground. 

External backgrounds are reduced to a negligible level thanks to the external shielding. \\
The veto rejection of events that are accompanied by an energy deposition above a chosen threshold (100 keV) in the liquid scintillator leads to a reduction of internal contribution by a factor of $\sim$3.5, resulting in an overall background level in the ROI of $\sim$0.2 cpd/kg/keV.

\section{Expected Sensitivity}\label{sec:sensitivity}

\begin{figure}[h!]
\centering
\includegraphics[scale=0.35]{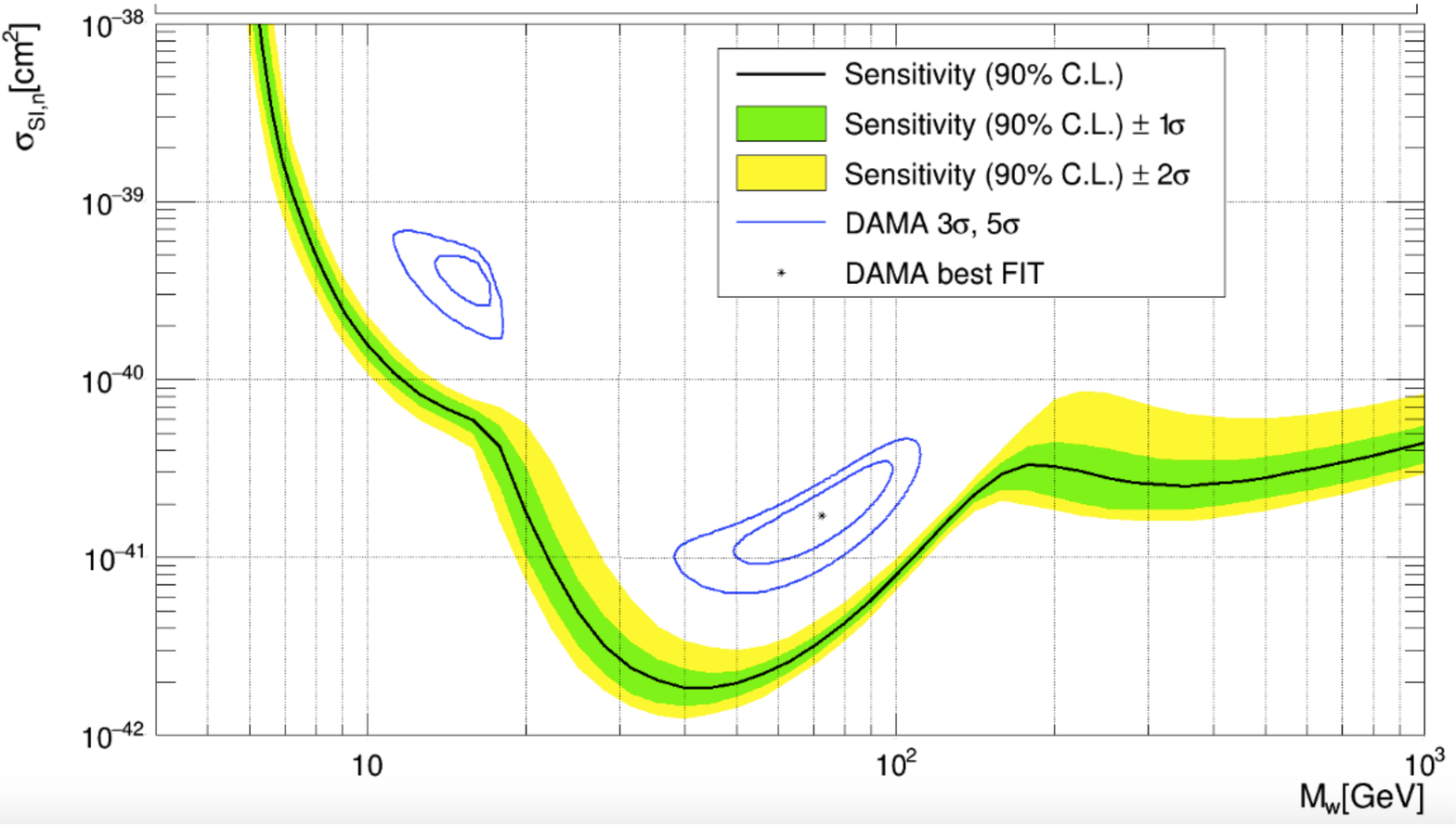}
\caption{90\% CL sensitivity plots for 3 years exposure time, 50 kg detector mass, recoil energy range 2-6 keV, background rate calculated with PoP Monte Carlo simulations.}
\label{sensi_plots}
\end{figure}

The SABRE sensitivity to the DM annual modulation has been evaluated in the ROI energy region assuming standard hypotheses for WIMPs and the halo model \cite{halo_model}. The total crystal mass considered is $\sim$50 kg and the data-taking time is 3 years. The background has been taken from the PoP Monte Carlo simulations. The nuclear recoil quenching factor of Na is taken from \cite{quenching_princeton} while the I quenching factor is taken from \cite{damappb}. The sensitivity at 90\% confidence level (CL) has been calculated and it is shown in Fig. \ref{sensi_plots}. The 1$\sigma$ and 2$\sigma$ bands have been obtained by varying the Na and I quenching factors, the energy resolution, the detection efficiency and the background within the uncertainties.\\This study highlights that SABRE will be able to confirm or exclude the DAMA modulation signal with a significance of at least 5$\sigma$ in 3 years and with a mass of 50 kg. 

\end{document}